\documentclass[twocolumn,showpacs,preprintnumbers,amsmath,amssymb]{revtex4}

\usepackage{graphicx}
\usepackage{dcolumn}
\usepackage{bm}

\begin{document}

\title{
Dissipation Properties of Coupled Cavity Arrays
}

\author{Ke Liu}
\author{Lei Tan\footnote{corresponding author}}\email{tanlei@lzu.edu.cn}
\affiliation{%
School of Physical Science and Technology, Lanzhou University, Lanzhou 730000,  China}%

\date{\today}

\begin{abstract}
We propose an approach to analyze the dissipation properties of
coupled cavity arrays. Employing a kind of quasi-boson, it is
shown that the coupling to a bath renormalizes the localized mode
and the interaction between cavities. By virtue of without having
to mention the coordinates of bath, this approach would be great
conceptual and, moreover, computation advantage. Based on the
result, a single-photon transport in the array is examined, and
the total transmission rate is presented. Besides, we also suggest
a parameter to scale quality of the array.
\end{abstract}

\maketitle

Coupled cavity arrays(CCAs), the effective and manipulatable
many-body system~[\onlinecite{CCA}], are known playing a key role
in quantum information and quantum devices
progressing~\cite{ID1,ID2}, mimicking and studying strong
correlated physics~[\onlinecite{SC1,SC2,SC3}]. In the term ``CCA",
``cavity" refers to a region of space within which photon can be
efficiently confined, and ``coupled" usually implies individual
cavities resonantly coupled to each other via the evanescent
field. The boost of experimental techniques and advances in
theories, over the past years, have paved the way for researching
CCA systems more in-depth. CCAs are not only explored in various
forms, but also manufactured with smaller mode volume, higher
quality factor, more accurate addressability, and increased
number, of cavities lying in the set-ups~\cite{EX1,EX2}. Moreover,
the theoretical proposals, paradigmatically like tight-binding
model(TBM)~\cite{TBM1,TBM2}, Bose-Hubbard
model(BHM)~\cite{SC2,SC3,BHM}, and effective spin
model~\cite{Spin1,Spin2}, have been put forwarded and predicted a
large quantity of novel
applications~[\onlinecite{app1,app2,app3}].

Despite the substantial progress, both in experiment and theory,
still many important issues are unsolved. A problem recurs often
and in more than one facet is how dissipation would behave in CCA
systems~\cite{SC3,decay}. It is always referred to a system,
having numerous degrees offreedom, interacts with a bath, with
many, in principle infinitely many, modes. The huge Hilbert space
results in great hurdle, even challenge, to fully describe the
properties of such system, as yet, there is still lack a method to
understand and calculate properly. However, the surge of interest
in, so called, quantum manipulation and quantum simulation makes
the problem becoming urgent to be solved.

In this paper, we address the issue and show that, for weakly
intercavity coupled high-$Q$ array, most experiments in CCAs are
carried out such that favour conditions where coupled-mode
approximation is valid~[\onlinecite{TBvalid}], the problem can be
cured by using a kind of quasi-boson picture. Which essence is
discarding the coordinates of bath, and describing the system via
a effective Hamiltonian. Throughout our discussion, we base on the
vacuum, realized, and most general 1D CCA. The reasons are as
follows: Firstly, photon leaking from cavity mode is one of the
main process of dissipation. Secondly, the approach could also
account other sorts of dissipation, like atomic decay, in CCA
systems. Finally, and most importantly, it maintains the
universality of the proposed approach. According to examine the
single photon transport in CCA, the theoretical results are in
well agreement with experiments. Which indicates that our work may
open up wealth of possibilities for studying dissipative CCA
systems conveniently.
\begin{figure}[b]
\includegraphics[scale=0.95]{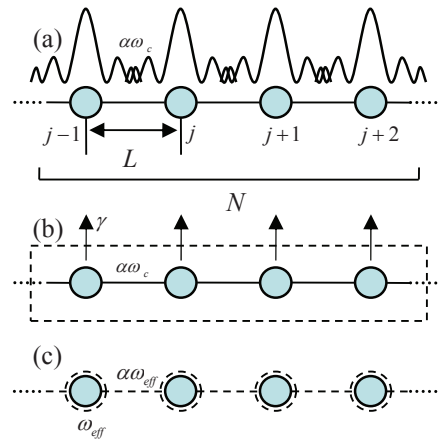}
\caption{\label{fig:epsart}Schematics of CCA.(a) Individual
cavities resonantly coupled to each other due to the overlap of
their evanescent fields of nearest neighbors.(b) The coupling of
CCA to a bath.Each resonator has a leakage rate $\gamma$.(c)
Effective treatment in quasi-boson picture,where system can be
regarded as a chain of quasi-bosons.}
\end{figure}

We are starting by reviewing the configuration of ideal CCA. As
shown in Fig.1(a), $N$ cavities, with a single mode characters by
frequency $\omega_c$, are arranged in a period $L$. The coupling
parameter in CCA is mathematically described by a specific overlap
integral $\alpha$. Because individual modes are confined
efficiently, only modes of nearest neighbor cavities have a small,
but non-vanishing, overlap, denoted by $\alpha\omega_c$. Such
configuration is well understood by using TB Scheme and forms the
basis for other CCA systems~[\onlinecite{CCA}]. In the Coulomb
gauge, these localized modes, labelled by $\varphi_j$, obey the
Maxwell equation~\cite{TBM1,Glauber}:
\begin{equation}
\frac{\epsilon(\bf
r)\omega^2_c}{c^2}\varphi_j-\nabla\times(\nabla\times\varphi_j)=0,
\end{equation}
and $\alpha$ is given by
\begin{equation}
\alpha=\int dr[\epsilon_{0}(\bf r)-\epsilon(\bf
r)]\varphi^\ast_{j}\varphi_{j+1}.
\end{equation}
Where $\bf r$ is a given, in fact, three dimensional vector,
$\epsilon_{0}(\bf r)$ the dielectric constant of single cavity,
and $\epsilon(\bf r)$ the periodic dielectric constant of array.

When $N$ is large enough, periodic boundary condition of CCA is
fulfilled and the Hamiltonian (with $\hbar=1$) reads
\begin{equation}
H_{array}=\omega_{c}\sum_{j}c^{+}_{j}c_{j}-\alpha\omega_{c}\sum_{\langle
j,j^{\prime}\rangle}c^{+}_{j}c_{j^{\prime}}.
\end{equation}
The bosonic operator $c^{+}_{j}$($c_{j}$) creates(annihilates) a
excited state at $j$th cavity, $\sum_{\langle
j,j^{\prime}\rangle}$ sums all pairs of cavities which are nearest
neighbors.

As a result of TB interaction, the whole system is no longer a
monochromatic field but splits into $N$ resonant modes, which are
well explained as liner combination of individual cavity modes,
and forms a narrow band in vicinity of $\omega_c$. The spectrum
takes the form
\begin{equation}
\omega(k)=\omega_{c}+2\alpha\omega_{c}\cos{k_{n}L}
\end{equation}
and has been observed experimentally by measuring the
transmission-phase properties~[\onlinecite{dispersion}]. However,
it is worth stressing that the wave vector,
$k_{n}=\frac{n\pi}{N+1}\frac{1}{L}$ for $n=1$ to $N$, does not has
a direct meaning in terms of photonic momentum but to analog of
the lattice vector in solid state physics.

Since we have not added any other features, like Jaynes-Commings
interaction and Kerr interaction, the formalism described above is
the most common foundation of CCA systems. Therefore, it is
advisable to analyze the dissipation properties base on the
scheme. In what follows, we will approach the problem in two
steps.

Firstly, we consider a single cavity coupled to a~bath composed of
the infinite set of harmonic oscillators~[\onlinecite{book1}]. The
bath generally has a continuous spectrum characterized by
$\omega_r$ and the density of states described by
$\rho(\omega_r)$. For simplicity, assuming here only one excited
state occupied by either cavity or bath, and the corresponding
probability amplitude denoted by $e_c$ and $e_r$ respectively.
Besides, because the totality is conservative, one can write the
eigenvalue equation as
\begin{subequations}
\label{eq:whole}
\begin{equation}
H|\varphi\rangle=\omega|\varphi\rangle,
\end{equation}
with
\begin{equation}
H=\omega_{c}c_{0}^{+}c_{0}+\int d\omega_{r}r^{+}r+\int
d\omega_{r}[\eta^{\ast}(\omega_{r})r^{+}c_{0}+h.c.],
\end{equation}
\begin{equation}
|\varphi\rangle=e_{c}c_{0}^{+}|\emptyset\rangle+\int
d\omega_{r}\rho(\omega_{r})e_{r}r^{+}|\emptyset\rangle.
\end{equation}
\end{subequations}
$r^{+}(r)$, which satisfies the commutation relation
$[r(\omega_{r}),r^{+}(\omega^{\prime}_{r})]=\delta(\omega_{r}-\omega^{\prime}_{r})$,
creates(destroys) an excited state of bath. $\eta(\omega_{r})$
represents the coupling strength between the two, $\omega$ the
total energy, and $|\emptyset\rangle$ the vacuum state.

Taking the inner product first with $c_{0}^{+}|\emptyset\rangle$
and then with $r^{+}|\emptyset\rangle$ to Eq.(5a),
\begin{subequations}
\label{eq:whole}
\begin{equation}
\omega_{c}e_{c}+\int
d\omega_{r}\rho(\omega_{r})\eta(\omega_{r})e_{r}=\omega e_{c},
\end{equation}
\begin{equation}
\omega_{r}e_{r}+\eta^{\ast}(\omega_{r})e_{c}=\omega e_{r}.
\end{equation}
\end{subequations}
From Eq.(6b),
$e_{r}=\frac{\eta^{\ast}(\omega_{r})}{\omega-\omega_{r}}e_{c}$,
and plugging it into Eq.(6a),
\begin{equation}
\omega_{c}e_{c}+\int
d\omega_{r}\rho(\omega_{r})\frac{|\eta(\omega_{r})|^{2}}{\omega-\omega_{r}}e_{c}=\omega
e_{c}.
\end{equation}
Note that
\begin{eqnarray}
&&\int d\omega_{r}\rho(\omega_{r})\frac{|\eta(\omega_{r})|^{2}}{\omega-\omega_{r}}\nonumber\\
&&=\int d\omega_{r}\rho(\omega_{r})\frac{|\eta(\omega_{r})|^{2}}{\omega-\omega_{r}+i\delta}\nonumber\\
&&=P\int
d\omega_{r}\rho(\omega_{r})\frac{|\eta(\omega_{r})|^{2}}{\omega-\omega_{r}}-i\pi\rho(\omega)|\eta(\omega)|^{2}
\end{eqnarray}
In the above derivation, we have extended the integration into
complex plane and used the relation
$\lim\limits_{y\!\rightarrow\!0^{+}}\frac{1}{x+iy}=~P\frac{1}{x}-i\pi\delta(x)$,
where $P$ denotes the Cauchy principal value, $x$ and $y$ are real
variables.

It is reasonable to assume that $\omega$ can cause excitation of
the cavity sharply peaks around $\omega_{c}$. Therefore, we can
evaluate Eq.(8) at $\omega=\omega_{c}$,
\begin{subequations}
\label{eq:whole}
\begin{equation}
P\int
d\omega_{r}\rho(\omega_{r})\frac{|\eta(\omega_{r})|^{2}}{\omega-\omega_{r}}
\approx P\int
d\omega_{r}\rho(\omega_{r})\frac{|\eta(\omega_{r})|^{2}}{\omega_{c}-\omega_{r}}
=\delta\omega_{c},
\end{equation}
\begin{equation}
i\pi\rho(\omega)|\eta(\omega)|^{2}\approx
i\pi\rho(\omega_{c})|\eta(\omega_{c})|^{2}=i\gamma.
\end{equation}
\end{subequations}
$\delta\omega_{c}$ is known analogous to the Lamb shift and
significantly small in the case of coupling to surroundings
weakly. $\gamma$ is the decay rate, which indicates a finite
lifetime of cavity mode.

Thus Eq.(7) becomes
\begin{equation}
(\omega_{c}+\delta\omega_{c}-i\gamma)e_{c}=\omega e_{c}.
\end{equation}
Which means that due to the coupling, the cavity mode is
renormalized by reckoning in frequency shift and intrinsic loss.
The above expression is equivalent to Eqs.(6a) and (6b), however,
does not contain degrees of freedom of bath. It motivates us to
introduce a quasi-boson described by $b$ and having a complex
eigenfrequency $\omega_{eff}=~\omega_{c}-~i\gamma$, where
$\delta\omega_{c}$ has been absorbed into $\omega_{c}$, to
redescribe the cavity mode. And then rephrasing Eqs.(5a)-(5c),
\begin{equation}
H_{eff}|\varphi\rangle=\omega_{eff}|\varphi\rangle,
\end{equation}
with the effective Hamiltonian $H_{eff}=\omega_{eff}b^{+}b$ and
now $|\varphi\rangle=e_{c}b^{+}|\emptyset\rangle$ referred to as
quasinormal-mode~[\onlinecite{QNM}]. Because of loss energy, the
system would nonconservative, and the corresponding operators
non-Hermitian. To compare the two descriptions, the communication
relation of $b$ reads $[b,b^{+}]=1+i\frac{2\gamma}{\omega_{c}}$.
Clearly, $\frac{2\gamma}{\omega_{c}}$ in order of $\frac{1}{Q}$,
thus bosonic communication relation is approximately satisfied.

Then next, we return to the case, see Fig.1(b), CCA coupled to a
bath and each resonator has a leakage rate $\gamma$. It is
verified experimentally the main sources of loss are individual
cavities, while the additional loss caused by periodic structure
is negligible~[\onlinecite{EX2}]. Combining the characteristic
cavities are weakly coupled, such system can be regarded as a
chain of quasi-bosons, see Fig.1(c), and mapped safety onto TB
scheme. According to Eq.(1), the associated eigenmodes, labelled
by $\psi_{j}$, satisfy
\begin{equation}
\frac{\epsilon(\bf
r)}{c^2}(\omega^{2}_{c}+\gamma^{2})\psi_j-\nabla\times(\nabla\times\psi_j)=0.
\end{equation}
For $\gamma^{2}$ is $Q^{2}$ orders of magnitude smaller than
$\omega^{2}_{c}$, the minimal loss on each lattice site does not
generate noticeable alteration to localized modes. Which also
illustrates that quasi-boson picture is an excellent approximation
to the established mode.

Consequently, the relevant overlap integral, $\alpha^{\prime}$, is
given by
\begin{eqnarray}
&&\alpha^{\prime}=\int dr[\epsilon_{0}(\bf r)-\epsilon(\bf r)]\psi^\ast_{j}\psi_{j+1}\nonumber\\
&&\quad\approx \int dr[\epsilon_{0}(\bf r)-\epsilon(\bf r)]\varphi^\ast_{j}\varphi_{j+1}\nonumber\\
&&\quad=\alpha.
\end{eqnarray}

Hence, we reach the familiar Hamiltonian but take dissipation into
account,
\begin{equation}
H=\omega_{eff}\sum_{j}b^{+}_{j}b_{j}-\alpha\omega_{eff}\sum_{\langle
j,j^{\prime}\rangle}b^{+}_{j}b_{j^{\prime}}.
\end{equation}
Yet interestingly, without having to mention the external degrees
of freedom, the effective treatment would be of great conceptual
and, moreover, computational advantage rather than treatment of
universe~\cite{decay}. One key feature is now the loss seems owing
to the nonideal boundary but not field oscillation, viz described
by a constant but not operators. In addition, it should also to
point out that the specific impact of renormalization to
interaction terms may vary from case to case, nevertheless all of
those represented by a small quantity $\alpha\gamma$. This is
consistent with the conditions discussed previous.

To demonstrate the validity of our approach, we consider now the
single-photon transport in the CCA. In the simplest possible
context, we assume that a photon has somehow been inject into 1st
cavity and propagating to the right. The frequency, $\omega$, of
photon satisfy dispersion relation (4), thus photon hopping can
occur between neighboring cavities due to the overlap of the light
modes. So the problem we treated can be described by Hamiltonian
(14). Furthermore, to focus on the total transmission rate, we can
restrict us to solve the stationary Sch\"{o}dinger equation
\begin{equation}
H|\psi\rangle=\omega|\psi\rangle,
\end{equation}
with ~$\psi=\sum\limits_{j}e_{j}b^{+}_{j}|\emptyset\rangle$, and
take~\cite{ej1,ej2} ~$e_{j}=~\left\{\begin{array}{l}
e_{j^{-}}=e^{ik_{n}sL}+r_{j}e^{-ik_{n}sL}\quad s<j\\
e_{j^{+}}=t_{j}e^{ik_{n}sL}\quad s>j\end{array}\right.$, $s=1$ to
~$N$. Where $r_{j}$ and $t_{j}$ denote the local transmission
amplitude and  reflection amplitude of photon respectively.
Solving Eq.(15) by using the continuous condition
$e_{j^{-}}=e_{j^{+}}$ at $j$th site and the constraint
condition$|r_{j}|^{2}+|t_{j}|^{2}\leq1$ due to the irreversible
loss of energy, we get
\begin{subequations}
\label{eq:whole}
\begin{equation}
r_{j}=\frac{\kappa\cos{k_{n}L}-\gamma}
{(\gamma+\xi|\sin{k_{n}L}|-\kappa\cos{k_{n}L})-i\kappa|\sin{k_{n}L}|}
e^{i2k_{n}jL},
\end{equation}
\begin{equation}
t_{j}=\frac{(\xi-i\kappa)|\sin{k_{n}L}|}
{(\gamma+\xi|\sin{k_{n}L}|-\kappa\cos{k_{n}L})-i\kappa|\sin{k_{n}L}|}.
\end{equation}
\end{subequations}
Above, $\xi=2\alpha\omega_{c}$ and $\kappa=2\alpha\gamma$ for
compactness, $e^{i2k_{n}jL}$ is position-dependent global phase
but does not affect the transport properties, and the absolute
value sign is need for energy conservation.

Before proceeding, here we briefly outline some of the main
features of $r_{j}$ and $t_{j}$. The nonzero reflection amplitude
is caused by local loss. Under the circumstance of system is
confined in one dimension, incoming photon having possibility to
escape toward the opposite direction. Note however, this
possibility would not make photon enters the previous cavity and
becomes left-moving photon, but eventually decay to other
dimensions. Local loss also leads to the nonunitary transmission
amplitude. By dropping the second-order small quantity $\kappa$,
the maximum of transmission coefficient approximates
$\frac{1}{(1+\gamma/\xi)^{2}}$, which means the local transport
properties is determined by the competition between photon hopping
and decay, since they are the only channels photon can leave a
certain cavity.

And then, the total transmission rate, $T$, can be intuitively
written as
\begin{equation}
T=\prod_{j}|t_{j}|^{2}=|t_{j}|^{2N}.
\end{equation}
\begin{figure}

\includegraphics[scale=0.85]{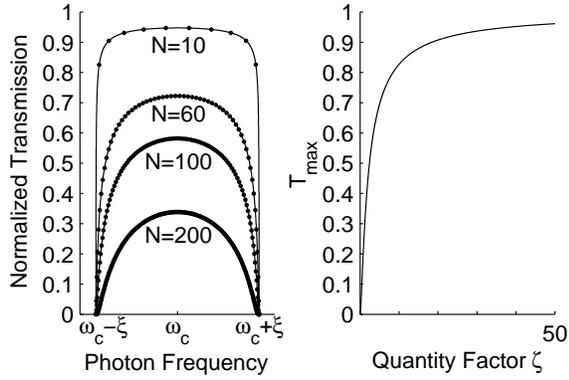}
\caption{\label{fig:epsart}(a) Transmission spectrum for
single-photon transport in CCA. The total transmission rate of
each splitted resonant mode is denoted by dots and fitted by a
curve. Following the experimental parameters in Ref.~\cite{EX2},
$Q=1.1\times10^{6}$ and $\xi=6.47\times10^{-4}\omega_{c}$, the
output power, concretely like when $N=60$, is well agreement. (b)
Dependence of the maximal transmission rate on quantity factor
$\zeta$. In Fig.(2a), $\zeta$ equal to $71.17$, $11.86$, $7.117$,
and $3.559$ respectively.}
\end{figure}

The transmission spectrum, shown in Fig.2(a), retains the symmetry
of dispersion relation (4) and vanishes at band edges. When the
propagating photon is on resonant with individual cavity,
$\omega=\omega_c$, the spectrum exhibits the maximum,
$T_{max}\approx\frac{1}{(1+\gamma/\xi)^{2N}}$. While the ratio
between local loss rate and intercavity coupling strength is far
less than one, we can take
$(1+\frac{\gamma}{\xi})^{2N}=1+\frac{2N\gamma}{\xi}+\cdots$, thus
\begin{equation}
T_{max}\approx\frac{1}{(1+N\gamma/\xi)^{2}}=\frac{1}{(1+N/\alpha
Q)^{2}}.
\end{equation}

By substituting $Q=\frac{2\omega_{c}}{\gamma}$ and
$\xi=2\alpha\omega_{c}$, the maximal transmission rate now is
described directly with three essential parameters of CCA. It is
helpful to define a new quality factor, $\zeta=\frac{\alpha
Q}{N}$, to scale CCA's transport properties, which lead to
\begin{equation}
T_{max}=\frac{1}{(1+\zeta)^{2}}.
\end{equation}
Furthermore, the transport loss mainly stems from cavity-mode
decay, thus $\zeta$ could reflect as well as dissipation
properties for other CCA systems. A high-$\zeta$ array, see
Fig.1(a) for example, is often referred to steep or sharp
spectrums.

In summary, to aim at the descriptive difficulty caused by the
coupling of CCA systems to environment, we have proposed a kind of
quasi-boson picture and shown its effectiveness by analyzing the
single-photon transport. Here we would like to emphasize the
generality of our approach, which is capable of treating dynamical
problems and other sorts of dissipation~\cite{ej2,ohmic} and
provides a starting point for discussing more complicated
situations~[\onlinecite{CCA}].

This work was partly supported by the National Natural Science
Foundation of China under Grant No. $10704031$, the National
Natural Science Foundation of China for Fostering Talents in Basic
Research under Grant No.$J0730314$, and the Natural Science
Foundation of Gansu Under Grant No. 3ZS061-A25-035.

\end{document}